\documentclass[%
 reprint,
 superscriptaddress,
 showpacs,
 showkeys,
 amsmath,amssymb,
 aps,
 prd,
 lengthcheck,%
]{revtex4}
\usepackage{bbding}

\usepackage{graphicx}
\usepackage{dcolumn}
\usepackage{bm}
\usepackage{hyperref}

\usepackage{color}
 \newcommand{\be}{\begin{equation}}
 \newcommand{\ee}{\end{equation}}
 \newcommand{\ba}{\begin{eqnarray}}
 \newcommand{\ea}{\end{eqnarray}}
 \newcommand {\calG} {{\mathcal G}}
 \newcommand {\calT} {{\mathcal T}}

 \newcommand {\calY} {{\mathcal Y}}

\def \be {\begin{equation}}
\def \ee {\end{equation}}
\def \ba {\begin{eqnarray}}
\def \ea {\end{eqnarray}}
\newcommand {\dbar}{{d\kern-.22em\lower-.73ex\hbox{-}}}

\newcommand {\bP} {{\mathbf P}}
\newcommand {\bp} {{\mathbf p}}

\newcommand {\bn} {{\mathbf n}}
\newcommand {\brr} {{\mathbf r}}

\newcommand {\calM}{{\mathcal M}}

\newcommand {\calO}{{\mathcal O}}

\begin{document}


\title{More on volume dependence of spectral weight functions}
\thanks{Work supported in part by NSFC under grant No.10835002 and No.10721063.}%

\author{Zhi-Yuan Niu}
\affiliation{%
School of Physics, Peking University, Beijing 100871, P.~R.~China
}%


\author{De-Chuan Du}
\affiliation{%
School of Physics, Peking University, Beijing 100871, P.~R.~China
}%

\author{Bao-Zhong Guo}
\affiliation{%
School of Physics, Peking University, Beijing 100871, P.~R.~China
}%

\author{Ning Li}
\affiliation{%
School of Physics, Peking University, Beijing 100871, P.~R.~China
}%

\author{Chuan Liu}%
\email{liuchuan@pku.edu.cn}
\affiliation{%
School of Physics and Center for High Energy Physics, Peking
University, Beijing 100871, P.~R.~China
}%

\author{Hang Liu}
\affiliation{%
School of Physics, Peking University, Beijing 100871, P.~R.~China
}%






\date{\today}

\begin{abstract}
 Spectral weight functions are easily obtained from
 two-point correlation functions
 and they might be used to distinguish single-particle from
 multi-particle states in a finite-volume lattice calculation, a problem crucial for
 many lattice QCD simulations.
 In previous studies, it is shown that the spectral weight function
 for a broad resonance shares the typical volume dependence of
 a two-particle scattering state i.e. proportional to $1/L^3$ in
 a large cubic box of size $L$ while the narrow resonance case
 requires further investigation. In this paper, a generalized formula
 is found for the spectral weight function which incorporates both
 narrow and broad resonance cases. Within L\"uscher's formalism,
 it is shown that the volume dependence of
 the spectral weight function exhibits a single-particle behavior
 for a extremely narrow resonance and a two-particle behavior
 for a broad resonance. The corresponding formulas for both $A^+_1$
 and $T^-_1$ channels are derived. The potential application of
 these formulas in the extraction of resonance parameters are also
 discussed.
\end{abstract}

\maketitle


 \section{\label{sec:intro}Introduction}

 Lattice Chromodynamics (lattice QCD) has provided us with
 a genuine non-perturbative theoretical framework to study low-energy
 physics in strong interactions from first principles.
 Combined with numerical Monte Carlo simulations, lattice QCD has become
 a crucial player in hadron spectroscopy and other studies
 in the field.

 In a typical lattice simulation, one computes a two-point function of an appropriate operator
 with a definite quantum number via Monte Carlo simulation
 and the energy eigenvalues in that sector are then extracted
 from the time-dependence of the correlation function.
 In fact, the correlation function of an operator $\calO$
 may be written schematically as:
 $C(t)=\langle\calO(t)\calO(0)\rangle=\sum_n W_n(E_n)e^{-E_nt}$
 where $E_n$'s are the energy eigenvalues of the QCD Hamiltonian
 for the corresponding state $n$ with the quantum numbers associated with the operator
 $\calO$. The functions $W_n(E_n)$ are the so-called spectral weight
 functions for the state $n$.
 Since lattice simulations are all performed in a finite volume,
 the eigenvalues $E_n$ are all discrete in nature. This poses a
 question on the particle nature of the corresponding state.
 If in the infinite volume, single-particle and multi-particle
 states are different in the sense that the spectra for
 single particle states are discrete while that
 for multi-particle states are continuous starting from
 the multi-particle threshold. In a finite volume,
 however, it is not an easy task to tell whether a state
 with a particular energy $E_n$ is a single- or a multi-particle state
 since all states have become discrete. This distinction
 is not only for academic purposes but also of phenomenological
 importance as well if one recalls
 that quite a number of new resonance-like structures have been identified
 rather close to the multi-particle threshold of known hadrons.
 To cope with this problem, it has been suggested that the
 volume dependence of the spectral weight functions $W_n(E_n, L)$, which is
 also a function of the box size $L$, may be used
 as a probe to distinguish single- and multi-particle
 states~\cite{KFLiu04:pentaquark}. With an appropriate normalization, it is
 argued that the spectral weight function for
 a single-particle state has little volume dependence
 while that for a two-particle state will exhibit
 a $1/L^3$ behavior.

 In our previous study~\cite{Niu:2009gt}, it was shown that the above mentioned
 criteria becomes invalid for a broad resonance. Instead of
 showing little volume dependence, the spectral weight function
 for a broad resonance will exhibit a $1/L^3$ behavior, typical
 for  two-particle states. However, our previous study does not
 apply for extremely narrow resonances.
 In this paper, we re-examine the problem within
 L\"uscher's formalism (see Ref.~\cite{luscher91:finitea}),
 with an emphasis on the narrow resonance limit case.
 A generalized formula is found which
 incorporates both narrow and broad resonance cases.
 The conclusion we reached is exactly analogous to what
 we discovered in a model study~\cite{chuan08:Lee_model},
 namely that for a broad resonance, the spectral weight behaves
 like a multi-particle state while for an extremely narrow
 resonance, a single-particle behavior realizes.
 We also point out the possibility to extract the resonance parameters
 from the spectral weight function.
 We closely follow the discussion of our previous
 work~\cite{Niu:2009gt}. The reader is referred to that
 reference for notations and further details.

 \section{\label{sec:model}The model in the infinite volume and the signature of a resonance}

 In the infinite volume, consider a non-relativistic
 quantum mechanical model with Hamiltonian
 given by:
 \be
 \label{eq:hamiltonian}
 H= -{1\over 2m}\nabla^2+V(r)\;,
 \ee
 where the potential $V(\brr)$ is zero for $r>a$ with some $a>0$.
 We now discuss the energy eigenstates satisfying:
 $H\Psi(\brr)=E\Psi(\brr)$.
 One can expand the eigenfunction as:
 $\Psi(\brr)=\psi_{lm}(r)Y_{lm}(\bn)$
 with: $\brr=r\bn$ and $\psi_{lm}(r)$ is
 the radial wave-function satisfying the radial Schr\"odinger
 equation:
 \be
 \label{eq:radial}
 \left({d^2\over dr^2}+{2\over r}{d\over dr}-{l(l+1)\over r^2}
 +k^2-2mV(r)\right)\psi_{lm}(r)=0\;.
 \ee
 where $E=k^2/(2m)$ being the energy eigenvalue of the state.
 There exists a unique solution to the radial
 Schr\"odinger equation that is bounded near the origin
 which is denoted as: $u_l(r;k)$
 and the general solution to the radial Schr\"odinger equation
 has the form: $\psi_{lm}(r)=b_{lm}u_l(r;k)$
 with some constant $b_{lm}$ to be fixed by other conditions.

 In the region $r>a$ where the interaction vanishes, the
 solution $u_l(r;k)$ are expanded in terms of spherical
 Bessel functions:
 \footnote{Here we adopt a convention for spherical Bessel
 functions as in Ref.~\cite{luscher91:finitea}.}
 \be
 \label{eq:ul_expand}
 u_l(r;k) = \alpha_l(k)j_l(kr) +\beta_l(k)n_l(kr)\;.
 \ee
 The coefficients $\alpha_l(k)$ and $\beta_l(k)$ have
 simple relation with the scattering phase shift:
 \be
 e^{2i\delta_l(k)}={\alpha_l(k)+i\beta_l(k)\over
 \alpha_l(k)-i\beta_l(k)}\;,
 \;\;
 \cot\delta_l(k)={\alpha_l(k)\over\beta_l(k)}\;.
 \ee
 In the low-energy limit $k\rightarrow 0$, one normally defines:
 \be
 \alpha^0_l=\lim_{k\rightarrow 0} k^l\alpha_l(k)\;,
 \;\;
 \beta^0_l=\lim_{k\rightarrow 0} k^{-l-1}\beta_l(k)\;,
 \ee
 and the threshold parameters $a_l\equiv \beta^0_l/ \alpha^0_l$.
 In particular, $a_0$ for $l=0$ is referred to as
 the $s$-wave scattering length.
 The threshold parameters $a_l$ are important because they
 characterize most behaviors in low-energy scattering processes.
 For example, we have:
 \be
 \delta_l(k) \simeq a_l k^{2l+1}+O(k^{2l+3})\;,
 \;\;
 (\mbox{\rm mod    } \pi)\;.
 \ee
 To fix the normalization of $u_l(r;k)$, we impose the condition:
 \footnote{Note that this condition is somewhat different
 from the original condition imposed by L\"uscher. We choose
 this condition because with this convention, the relation
 in Eq.~(\ref{eq:lueders}) is in a simpler form.}
 \be
 \label{eq:normalize_u}
 u_l(r;k) \simeq {1\over kr}\sin\left(kr-l\pi/2+\delta_l\right)\;.
 \ee
 Another way of writing the asymptotic behavior of the radial
 wave function in Eq.~(\ref{eq:ul_expand}) is as follows:
 \be
 u_l(r;k)\simeq {1\over r}\left[
 A_l(k)e^{ikr}+B_l(k)e^{-ikr}
 \right]\;,
 \ee
 where the functions $A_l(k)$ and $B_l(k)$ are
 related to the coefficients $\alpha_l(k)$ and
 $\beta_l(k)$ via
 \be
 A_l(k)=(-i)^l(\beta_l-i\alpha_l)/(2k),
 B_l(k)=(+i)^l(\beta_l+i\alpha_l)/(2k).
 \ee
 When regarded as functions of the
 energy $E=k^2/(2m)$, we will write them as $A_l(E)$ and
 $B_l(E)$, respectively.
 Note that for real positive energy, we have:
 \be
 A_l(E)=B_l(E)^*\;,
 \;\; E>0,E\in \mathbb{R}\;.
 \ee

 The condition for a narrow resonance is that,
 the coefficient $B_l(E)$ has a complex zero
 on the second Riemann sheet at $E=E_\star-i\Gamma/2$,
 close to the real axis:
 \be
 B_l\left(E_\star-i{\Gamma\over 2}\right)=0\;,
 \ee
 with $\Gamma$ being the physical width of the resonance.
 In this case, the scattering properties of the system
 is mainly governed by the position of this resonance pole,
 if the scattering energy $E$ is very close to the
 resonance energy $E_\star$.
 In fact, since the function $B_l(E)$ vanishes at
 $E_\star-i\Gamma/2$, for real energy $E$ close
 to $E_\star$, we may approximate the function by:
 \be
 B_l(E)\simeq b_l(E-E_\star+i\Gamma/2)\;,
 \ee
 for some complex constant $b_l$. Therefore, for
 positive real energy $E$ close to $E_\star$, the
 radial wavefunction looks like
 \be
 \label{eq:u_expand}
 u_l(r;k)\simeq {1\over r}\left[
 b^*_l(E-E_\star-i\Gamma/2)e^{ikr}
 +h.c.
 \right]\;.
 \ee
 This then shows that the phase-shift near the
 resonance point is given by
 \be
 \label{eq:delta_resonance}
 e^{2i\delta_l(E)}=e^{2i\delta^{(\star)}_l}
 \left({E-E_\star-i\Gamma/2\over E-E_\star+i\Gamma/2}
 \right)\;,
 \ee
 where the quantity $\delta^{(\star)}_l$ is given by
 \be
 e^{2i\delta^{(\star)}_l}=(-1)^{l+1}(b^*_l/b_l)\;.
 \ee
 This is a constant phase (meaning that it does not depend on
 the energy substantially in the resonance region) which depends only on the
 position of the resonance. The energy dependent part
 of the phase shift is given solely by the
 factor in the big parenthesis in Eq.~(\ref{eq:delta_resonance}).
 It is seen that, for extremely narrow resonance, the phase shift jumps
 by $\pi$ when the energy passes through $E_\star$ within
 an energy range of order of several $\Gamma$.

 Another important point has become clear from the
 expansion in Eq.~(\ref{eq:u_expand}). If we set the
 energy $E$ to be exactly the resonance energy $E_\star$,
 the expression $u_l$ is vanishing if $\Gamma$ becomes
 infinitesimally small. This means that, to properly
 normalize the wave-function, therefore,
 the coefficients $b_l$ has to be of order $1/\Gamma$.

 \section{\label{sec:model_torus}The Model on a Torus}

 Now we enclose the system we discussed in the previous section
 in a large cubic box of size $L$ and impose the periodic boundary condition
 in all three spatial directions.
 The potential itself is also periodically continued to
 $V_L(\brr)=\sum_{\bn\in\mathbb{Z}^3}V(|\brr+\bn L|)$.
 We define the so-called ``outer region" as
 \be
 \Omega=\{\brr:\; |\brr+\bn L|>a,\;\; \mbox{for all } \bn\in\mathbb{Z}^3\}\;.
 \ee
 This is the region where the potential vanishes identically.
 We assume $L \gg 2a$ so that the outer region
 admits free spherical wave solutions (asymptotic states).
 Following the discussion in our previous paper,
 we may write the true eigenfunction as:
 \be
 \label{eq:eigenfunction_exact}
 \Psi(\brr;k)=\sum_{lm} b_{lm}u_{l}(r;k) Y_{lm}(\bn)\;.
 \ee
 where the coefficients $b_{lm}$ are to be determined by boundary
 conditions and normalization.

 In the outer region, the eigenfunctions can also be expanded
 in terms of singular periodic solutions (SPS) of Helmholtz equation.
 These are defined as
 \be
 \label{eq:glm_def}
 G_{lm}(\brr;k)\equiv \calY_{lm}(\nabla)G(\brr;k)\;,
 G(\brr;k)={1\over L^3}\sum_\bp{e^{i\bp\cdot\brr}\over \bp^2-k^2}
 \;.
 \ee
 with $\calY_{lm}(\brr)\equiv r^lY_{lm}(\Omega_\brr)$ being
 the harmonic polynomials. We may write
 \be
 \Psi(\brr;k)|_{\brr\in\Omega}=\sum_{lm}v_{lm}G_{lm}(\brr;k^2)\;.
 \ee
 In the meantime, the outer solution can also be expanded
 in terms of spherical harmonics and the spherical Bessel
 functions $j_l(kr)$, $n_l(kr)$
 and the matrix elements $\calM_{l'm';lm}(k^2_i)$.
 The explicit expressions can be found in Ref.~\cite{luscher91:finitea}.

 It is also convenient to introduce the
 cubic version for the functions $G_{lm}$ defined
 by Eq.~(\ref{eq:glm_def}):
 \be
 \label{eq:cubic_SPS}
 \calG^{(\Gamma)}_{l;i}(\brr;k)=
 \calY^{(\Gamma)}_{l;i}(\nabla)G(\brr;k)\;,
 \ee
 where $\calY^{(\Gamma)}_{l;i}(\brr)$ is the cubic
 harmonics for a particular represention $\Gamma$ of
 the cubic group. The explicit expressions are listed
 in Appendix~\ref{app:cubic} at the end of this paper.

 \subsection{L\"uscher's formula in $A^+_1$ and $T^-_1$ channels}

 In this subsection, we will derive L\"uscher's formula in
 the $A^+_1$ and $T^-_1$ channel (corresponding to $s$-wave and
 $p$-wave in the continuum, respectively). Using similar arguments and
 the cubic harmonics given in the appendix, it is
 easy to generalize the results to other channels.

 We first examine the $A^+_1$ sector which is the analogue of
 $s$-wave scattering. A good approximation for the $s$-wave dominated eigenfunction can
 be written as a superposition of $l=0$ and $l=4$ cubic
 harmonics with the $s$-wave component much larger than that
 of $g$-wave. So, we may write the eigen-function in $A^+_1$ sector as:
 \begin{widetext}
 \be
 \label{eq:Psi_A1}
 \Psi^{(A^+_1)}(\brr;k) = b_{0}u_0(r;k)Y^{(A^+_1)}_{0}(\Omega_\brr)
 +b_{4}u_4(r;k)Y^{(A^+_1)}_{4}(\Omega_\brr)+\cdots\;,
 \ee
 \end{widetext}
 with $|b_{4}|\ll |b_{0}|$ in the large volume limit.
 In the outer region where the interaction vanishes, using relation~(\ref{eq:ul_expand})
 we have
 \begin{widetext}
 \be
 \label{eq:psi_in_out}
 \Psi^{(A^+_1)}(\brr;k)|_{\brr\in\Omega} =
  b_{0}[\alpha_0j_0(kr)+\beta_0n_0(kr)]Y^{(A^+_1)}_{0}(\Omega_\brr)
 +b_{4}[\alpha_4j_4(kr)+\beta_4n_4(kr)]Y^{(A^+_1)}_4(\Omega_\brr)
 +\cdots\;.
 \ee
 \end{widetext}
 On the other hand,  in the outer region $\Omega$,
 we may also expand the wavefunction in terms of
 cubic SPS of Helmholtz equation defined in Eq.~(\ref{eq:cubic_SPS}):
 \begin{widetext}
 \be
 \label{eq:psi_into_Glm}
 \Psi^{(A^+_1)}(\brr;k)|_{r\in\Omega}= \left({4\pi\over k}\right)
 v_{0}\left[\calG^{(A^+_1)}_{0}(\brr;k)
 +{v_{4}\over k^4}\calG^{(A^+_1)}_{4}(\brr;k)
 +\cdots\right]\;.
 \ee
 \end{widetext}
 Matching the above two expansions in the outer region
 and using the explicit expressions for the cubic harmonics,
 one arrives at the following result (i.e. L\"uscher's formula)
 for the $A^+_1$ sector:
 \be
 (\cot\delta^{(0)}-m_{00})(\cot\delta^{(4)}-m_{44})
 =m_{40}m_{04}\;.
 \ee

 For the sector $T^-_1$, which is the cubic analogue of
 the vector channel, we have a similar expression:
 \begin{widetext}
 \be
 \label{eq:Psi_T1}
 \Psi^{(T^-_1)}_3(\brr;k) = b_{1}u_1(r;k)Y^{(T^-_1)}_{1;3}(\Omega_\brr)
 +b_{3}u_3(r;k)Y^{(T^-_1)}_{3;3}(\Omega_\brr)+\cdots\;,
 \ee
 \end{widetext}
 where we have only written out the third component of the
 vector wavefunction. We may also expand in terms of
 cubic SPS of Helmholtz equation:
 \begin{widetext}
 \be
 \label{eq:T1_into_Glm}
 \Psi^{(T^-_1)}_3(\brr;k)|_{r\in\Omega}= \left({4\pi\over k^2}\right)
 v_{1}\left[\calG^{(T^-_1)}_{1;3}(\brr;k)
 +{v_{3}\over k^2}\calG^{(T^-_1)}_{3;3}(\brr;k)
 +\cdots\right]\;.
 \ee
 \end{widetext}
 Using the explicit expressions for the cubic harmonics and
 cubic SPS to match the above two equations we obtain
 L\"uscher's formula in the $T^-_1$ sector:
 \be
 (\cot\delta^{(1)}-m_{11})(\cot\delta^{(3)}-m_{33})
 =m_{13}m_{31}\;.
 \ee
 Since numerically we have: $m_{11}=m_{00}$, therefore
 if higher momentum contaminations (i.e. $l\ge 4$ for $A^+_1$
 and $l\ge 3$ for $T^-_1$) are neglected, the formula
 in the $T^-_1$ channel has the
 same form as the one for $A_1$ channel.

 \subsection{Spectral weight function in $A^+_1$ sector}

 As we pointed out in our previous study~\cite{Niu:2009gt},
 the spectral weight function has to be normalized in
 a proper manner so that the single-particle states
 has a conventional behavior. For an interpolating
 operator $\calO(t)$, the state $\calO^\dagger(0)|0\rangle$
 will generally be a superposition of a tower of
 energy eigenstates: $\{|E\rangle\}$ with energy
 eigenvalue $E$. The spectral weight for a particular
 energy state $|E\rangle$ obtained from
 the two-point correlation function of $\calO(t)$
 is given by:
 \be
 W(E)=|O(E)|^2=|\langle E|\calO^\dagger(0)|0\rangle|^2
 \;,
 \ee
 where $O(E)=\langle E|\calO^\dagger(0)|0\rangle$ is
 the overlap of the exact energy eigenstate $|E\rangle$
 with the state $|\calO^\dagger(0)|0\rangle$.

 The exact energy eigenfunction $\langle\brr_1,\brr_2|E\rangle$
 only depends on the relative coordinate $\brr=\brr_2-\brr_1$
 and does not depend on the center-of-mass of the two particles.
 Therefore, in terms of the wavefunctions associated with
 the relative motion, the normalization conditions for the
 $A^+_1$ and $T^-_1$ channels become:
 \ba
 \label{eq:a1_normalize}
 \int_{\calT_3} d^3\brr |\Psi^{(A^+_1)}(\brr;k)|^2 &=&{1\over L^3}\;.
 \\
 \label{eq:t1_normalize}
 \int_{\calT_3} d^3\brr \left[\Psi^{(T^-_1)}_i(\brr;k)\right]^*
 \cdot\Psi^{(T^-_1)}_j(\brr;k)&=&{1\over L^3}\delta_{ij}\;,
 \ea
 where $\Psi^{(A^+_1)}(\brr;k)$ and $\Psi^{(T^-_1)}_j(\brr;k)$
 is the wavefunction for the $A^+_1$ and $T^-_1$ channel
 respectively.

 Let us first investigate the normalization condition in the $A^+_1$ channel.
 The integral over the torus in Eq.~(\ref{eq:a1_normalize})
 runs over two regions: the inner region $B=\{\brr: r\le a, \mod L\}$
 where the potential is present and the outer region $\Omega$ where
 it vanishes:
 \be
  \int_B d^3\brr |\Psi^{(A^+_1)}(\brr;k)|^2
  +\int_{\Omega} d^3\brr |\Psi^{(A^+_1)}(\brr;k)|^2
  = {1\over L^3}\;.
 \ee
 If we assume that the wavefunction in this sector is dominated by the
 $s$-wave component such that
 \ba
 \Psi^{(A^+_1)}(\brr;k)|_{\brr\in B} &\simeq& b_0u_0(r;k)Y_{00}(\bn)\;,
 \nonumber \\
 \Psi^{(A^+_1)}(\brr;k)|_{\brr\in \Omega}&\simeq& \left({4\pi\over k}\right)v_0G_{00}(\brr;k)\;,
 \ea
 Matching the above two wavefunctions at $r=a$ will fix the
 ratio $b_0/v_0$. Therefore, we may write: $|b_0|^2=\zeta|v_0|^2$
 with $\zeta$ being some positive constant with the dimension of
 $k^{-2}$. Therefore, we may express the normalization condition as
 \begin{widetext}
 \be
 |v_0|^2\left[\zeta\int^a_0r^2 |u_0(r;k)|^2 dr
  +\left({4\pi\over k}\right)^2\int_{\Omega} d^3\brr |G_{00}(\brr;k)|^2
  \right]= {1\over L^3}\;.
 \ee
 \end{widetext}
 If we drop the first term in the bracket in the above equation
 and approximate the second integral by an integral over the
 whole torus, we obtain exactly the same formula as we derived
 in our previous study (see Eq.~(38) in Ref.~\cite{Niu:2009gt}
 and the discussions about it).
 Normally, if there were no sharp resonance
 in the scattering, neglecting of the first term is justified since
 the second term is proportional $L^3$, much larger than any
 normal term in the large volume limit.
 However, when an extremely narrow resonance is present,
 this is {\em not} the case, as we remarked
 at the end of Sec.~\ref{sec:model}.

 The first term in the normalization condition involves
 the exact radial wavefunction in the interaction region
 which seems to depend on the details of the potential.
 However, although generally speaking
 we do not know the exact form of the
 inner wavefunction $u_0(r;k)$, the integral
 $\int^a_0r^2 |u_0(r;k)|^2 dr$ can be related
 to the corresponding phase shift via an exact formula:
 \footnote{See for example pp. 555 of Ref.~\cite{landau:quantum_mechanics_book}
 for a derivation of this formula.
 According to Ref.~\cite{landau:quantum_mechanics_book}, this formula
 was first obtained by L\"uders in 1955.}
 \be
 \label{eq:lueders}
 \int^a_0r^2 |u_0(r;k)|^2 dr=
 2\left(a+{d\delta_0(k)\over dk}\right)
 -{1\over k}\sin 2(ka+\delta_0)\;.
 \ee
 Now, if there exists an extremely narrow resonance at
 a particular energy $E_\star=k^2_\star/(2m)$, then close
 to the resoance we have $\cot\delta_0\simeq -(E-E_\star)/(\Gamma/2)$.
 This implies that:
 \be
 {d\delta_0(k)\over dk}={k\over m}{d\delta_0\over dE}
 \simeq {\Gamma/2\over (E_\star-E)^2+\Gamma^2/4}\;.
 \ee
 It is seen that, close to the resonance energy,
 the derivative $d\delta_0/dk$ is proportional
 to $1/\Gamma$ in the narrow resonance limit (i.e. when $E\simeq E_\star$
 and $\Gamma\rightarrow 0$).
 Therefore, in the narrow resonance limit,
 such a contribution can become comparable
 or even larger than the second term that is
 proportional to $L^3$ in a finite volume.
 All the other terms in
 the integral $\int^a_0r^2 |u_0(r;k)|^2 dr$ are bounded
 in both the narrow resonance and the large volume limit.
 Therefore, taking into account of this possibility, the normalization condition
 for the wavefunction is written as
 \begin{widetext}
 \be
 \label{eq:normal_A}
 |v_0|^2\left[\xi+{\zeta\Gamma\over (E_\star-E)^2+\Gamma^2/4}
  +\left({4\pi\over k}\right)^2\int_{\calT_3} d^3\brr |G_{00}(\brr;k)|^2
  \right]= {1\over L^3}\;,
 \ee
 \end{widetext}
 where $\xi$ and $\zeta>0$ are two constants
 that are regular in both the large volume and
 the narrow resonance limit.

 The integral in Eq.~(\ref{eq:normal_A}) has been computed
 in our previous paper with the result~\cite{Niu:2009gt}:
 \be
 \left({4\pi\over k}\right)^2\int_{\calT_3} d^3\brr |G_{00}(\brr;k)|^2
 ={4\pi\over k^2}F'(k^2)\;,
 \ee
 with the function $F(k^2)$ defined by
 \be
 F(k^2)\equiv {1\over L^3}\sum_\bp {f(\bp^2)\over \bp^2-k^2}\;.
 \ee
 In the large volume limit, the derivative of this function
 has the approximate form
 \be
 F'(k^2)\simeq {1\over 8\pi k}\cot\delta_0(k)
 +{k\over 4\Delta\bp^2}\csc^2\delta_0(k)\;.
 \ee
 Collecting everything together, we obtain
 the following normalization condition in the $A^+_1$ sector:
 \begin{widetext}
 \be
 \label{eq:v_0_final}
 |v_0|^2\left[\xi+{\zeta\Gamma\over (E_\star-E)^2+\Gamma^2/4}
  +{1\over 2 k^3}\cot\delta_0(k)
 +{\pi\over k\Delta\bp^2}\csc^2\delta_0(k)
  \right]= {1\over L^3}\;.
 \ee
 \end{widetext}
 As discussed above, the second term
 in the bracket in the above formula is singular in the narrow
 resonance limit ($O(1/\Gamma)$) while
 the fourth term is singular in the
 large volume limit ($O(L^3)$). All other terms
 are regular in both limits.

 Following similar steps as in our previous paper~\cite{Niu:2009gt},
 the spectral weight function in the $A^+_1$ channel
 now takes the following form:
 \begin{widetext}
 \be
 \label{eq:spectral_final}
 W^{(A^+_1)}(E)
 ={4\pi |\varphi_L(k^2)|^2/k^2\over
 \xi+{\zeta\Gamma\over (E_\star-E)^2+\Gamma^2/4}
 +{1\over 2k^3}\cot\delta_0(k)
 +{\pi \over k\Delta\bp^2}\csc^2\delta_0(k)
 }\;,
 \ee
 \end{widetext}
 where the function $\varphi_L(k^2)$ has been introduced
 in Ref.~\cite{Niu:2009gt} which has little volume dependence.
 The main difference between this formula and the one obtained
 in our previous paper is the appearance of the term
 $(\zeta\Gamma)/[(E_\star-E)^2+\Gamma^2/4]$
 in the denominator. This term becomes singular in the
 narrow resonance limit (proportional to $1/\Gamma$, as $E=E_\star$).
 If the resonance is not extremely narrow,
 this term is non-singular and can be absorbed into
 the constant term $\xi$ and we recover our old result:
 the resonance acquires a spectral weight function
 proportional to $1/L^3$, typical for two-particle states.
 However, if the resonance
 is extremely narrow such that:
 \be
 \zeta/\Gamma \gg { 1\over k \Delta \bp^2}\;,
 \ee
 the denominator is dominated by the term containing $\Gamma$
 and the whole spectral weight function has little volume
 dependence, typical for a single-particle state.
 It is also noted that, in this limit, the spectral
 weight function is proportional to $\Gamma$, the physical width
 of the resonance. This is expected since when the resonance becomes
 extremely narrow, the coupling of the resonance with
 the two-particle final states also becomes infinitesimally small
 (and is proportional to the width as it should).
 To summarize, Eq.~(\ref{eq:spectral_final}) is a generalization of our
 previous formula in the sense that it incorporates
 both the large volume and the narrow resonance limit.

 \subsection{Spectral weight function in $T^-_1$ sector}

 Following a similar treatment as in the $A^+_1$ channel,
 we can obtain the relevant formulas in $T^-_1$ channel.
 We construct the state using an interpolating operator
 of the form:
 \begin{widetext}
 \be
 |\Phi\rangle=\calO^\dagger(0)|0\rangle={1\over \sqrt{L^3}}\sum_\bP \tilde{\Phi}(\bP)P_z\left[
 \tilde{\pi}^\dagger_1(\bP,0)\tilde{\pi}^\dagger_2(-\bP,0)
 -\tilde{\pi}^\dagger_2(\bP,0)\tilde{\pi}^\dagger_1(-\bP,0)\right]
 |0\rangle\;,
 \ee
 \end{widetext}
 where we have only taken one component (the third component)
 of the wavefunctions for a vector state.
 After factorizing out the center of mass motion,
 the normalization condition expressed in terms of the
 wavefunction for the relative motion becomes
 \be
 \label{eq:normalize_t1}
  \int_B d^3\brr |\Psi^{(T^-_1)}_3(\brr;k)|^2
  +\int_{\Omega} d^3\brr |\Psi^{(T^-_1)}_3(\brr;k)|^2
  = {1\over L^3}\;,
 \ee
 where the first integral is over the interaction region
 while the second is over the outer region where interaction
 vanishes identically.
 For the $p$-wave dominant wavefunction, we have
 $\Psi^{(T^-_1)}_3(\brr;k)|_{\brr\in B} \simeq b_1u_1(r;k)Y_{10}(\Omega_\brr)$
 for the wavefunction in the interaction region and
 $\Psi^{(T^-_1)}_3(\brr;k)|_{\brr\in \Omega}\simeq
 \left({4\pi/{k^2}}\right)v_1G_{10}(\brr;k)$
 in the outer region.
 \footnote{For convenience, a factor $(4\pi)/k^2$ is scaled out
 in the channel as opposed to $(4\pi)/k$ in the $A^+_1$ channel.}
 Matching the two wavefunction at the boundary yields the
 relation $|b_1|^2=\zeta'|v_1|^2$ for some positive real constant $\zeta'>0$.
 The first integral in Eq.~(\ref{eq:normalize_t1}) can again
 be estimated in the same manner as that for $A^+_1$ sector:
 \be
 \int^a_0r^2 |u_1(r;k)|^2 dr=
 2\left(a+{d\delta_1(k)\over dk}\right)
 -{1\over k}\sin 2(ka+\delta_1)\;.
 \ee
 Assuming that there exists a resonance at $E=E_\star$ in this channel
 and for energies that are close to this resonance we have:
 $\cot\delta(E)\simeq -(E-E_\star)/(\Gamma/2)$.
 Following similar derivations, we arrive at the following normalization condition
 for an energy eigenstate whose energy is in the resonance region:
 \begin{widetext}
 \be
 |v_1|^2\left[\xi'+{\zeta'\Gamma\over (E_\star-E)^2+\Gamma^2/4}
  +{3\over 2 k^3}\cot\delta_1(k)
 +{\pi\over k\Delta\bp^2}\csc^2\delta_1(k)
  \right]= {1\over L^3}\;,
 \ee
 \end{widetext}
 where $\xi'$ is some real parameter which is regular in both
 the narrow resonance and large volume limit.
 With this normalization condition in the $T^-_1$ sector,
 the spectral weight function takes the following form:
 \begin{widetext}
 \be
 \label{eq:spectral_final_t1}
 W^{(T^-_1)}(E)
 ={16\pi |\alpha + k^2\varphi_L(k^2)|^2/(3k^4)\over
 \xi'+{\zeta'\Gamma\over (E_\star-E)^2+\Gamma^2/4}
 +{3\over 2k^3}\cot\delta_1(k)
 +{\pi \over k \Delta\bp^2}\csc^2\delta_1(k)
 }\;,
 \ee
 \end{widetext}
 where the constant $\alpha$ is defined by:
 \be
 \alpha \equiv{1\over L^3}\sum_\bP \tilde{\Phi}(\bP)\simeq{1  \over  (2\pi)^3}\int d^3\bP\tilde{\Phi}(\bP)\;,
 \ee
 Inspecting Eq.~(\ref{eq:spectral_final_t1}) we find that,
 the behavior in the $T^-_1$ sector is the same as
 in the $A^+_1$ sector qualitatively. The second term
 in the denominator is singular in the narrow resonance limit
 while the fourth term is singular in the large volume limit.
 Which of these two terms dominates the spectral weight function
 depends on the size of the following two quantities in the
 denominator: $\zeta'/\Gamma$
 and $1/(k\Delta\bp^2)$. For an extremely narrow
 resonance: $\zeta'/\Gamma\gg 1/(k\Delta\bp^2)$ and the
 spectral weight function is almost volume independent,
 a typical behavior for a single particle state.
 On the other hand, if we have $\zeta'/\Gamma\ll 1/(k\Delta\bp^2)$
 the spectral weight exhibits a $1/L^3$ behavior, typical
 for a two-particle state.

 \section{Conclusions}

 In this paper, the volume dependence of the spectral weight
 function is analyzed with the emphasis on the possible narrow
 resonances. Generalizing our previous results on this subject,
 we obtain a formula which incorporates both the large volume
 limit and the narrow resonance limit.
 This is summarized in Eq.~(\ref{eq:spectral_final}) for the
 $A^+_1$ channel and in Eq.~(\ref{eq:spectral_final_t1}) for the
 $T^-_1$ channel. If so desired, formulas for other channels
 can also be derived similarly
 using the cubic functions listed in the Appendix.
 These formulae for the spectral weight functions
 exhibit the following general feature:
 near a resonance the spectral weight shows single-particle or multi-particle behavior
 depending on whether the resonance is narrow or broad
 in the corresponding box. This is controlled by the size
 of the two quantities $\zeta/\Gamma$ and $1/(k\Delta \bp^2)$ that
 appear in the spectral weight function.
 If the former one dominates, a single particle feature (an almost
 volume independent spectral weight function) realizes;
 if the latter one dominates, a two-particle feature emerges with
 a $1/L^3$ dependence on the volume for the spectral weight function.
 Therefore, using the general L\"uscher's formalism, we have shown
 this general feature for the spectral weight function, just as
 we discovered it in a solvable model in Ref.~\cite{chuan08:Lee_model}.

 Furthermore, the spectral weight function has a general dependence
 on the volume of the box and the physical width
 of the resonance in the resonance region.
 As we pointed out in our previous study~\cite{Niu:2009gt},
 this opens up a possibility to
 extract the resonance parameters from the spectral weight
 functions obtained in Monte Carlo simulations.
 It is readily verified from Eq.~(\ref{eq:spectral_final})
 and Eq.~(\ref{eq:spectral_final_t1}) that, close to the
 energy of a resonance, both spectral weight functions
 satisfy the following functional form:
 \be
 {1\over W(E,L)}\simeq A(E)+{B(E)\Gamma\over (E_\star-E)^2+\Gamma^2/4}
 + {C(E)\over\Delta E}\;,
 \ee
 where $A(E)$, $B(E)$ and $C(E)$ are smooth functions
 of the energy in the resonance region and they
 do not contain explicit volume dependence.
 The explicit volume dependence only comes from the term
 containing $\Delta E$ and the information about
 the resonance (i.e. $E_\star$ and $\Gamma$)
 is encoded in the second term of the above equation.
 If one performs simulations on a series of volumes with the
 energies close to an underlying resonance and fits the spectral weight
 functions, it is in principle possible to extract the resonance parameters.
 Of course, the feasibility of this remains to be checked in
 realistic simulations.


\appendix

 \section{Cubic harmonics for various irreducible representations}
 \label{app:cubic}

 In this appendix, we list the cubic harmonics used in our calculation for
 various irreducible representations of the cubic group.
 We denote these cubic harmonics as $\calY^{(\Gamma)}_{l,i}(\brr)$
 where $\Gamma$ is the label for the irrep, $l$ is the corresponding
 angular momentum label (for the rotational group) and
 $i=1,\cdots,dim(\Gamma)$ denotes different basis functions
 of the irrep $\Gamma$. If $\Gamma$ were one-dimensional, the
 label $i$ will be dropped.
 The functions $\calY^{(\Gamma)}_{l;i}(\brr)$ defined here
 are homogeneous functions of order $l$. We will also use
 the notation:
 \be
 Y^{(\Gamma)}_{l;i}(\Omega_\brr)=
 r^{-l}\cdot \calY^{(\Gamma)}_{l;i}(\brr)
 \ee
 to designates the corresponding angular functions.
 It is easy to see that they are linear combinations
 of ordinary spherical harmonics.
 Note that these functions are orthogonal
 to each other but some of them are {\em not} properly normalized.
 The table for these harmonics can be found in Ref.~\cite{vonderLage:1947zz}.

  \begin{widetext}
 {\bf \HandRight\ The $A^+_1$ channel:}
 \be
 \label{eq:cubicY_A1}
 \left\{ \begin{aligned}
 \calY^{(A^+_1)}_{0}(\brr) &=\calY_{00}(\brr)\;, \\
 \calY^{(A^+_1)}_{4}(\brr) &=\calY_{40}+\sqrt{5\over
 14}(\calY_{4,4}+\calY_{4,-4})
 \propto \left(x^4+y^4+z^4-{3\over 5} r^4\right)\;,\\
 \calY^{(A^+_1)}_{6}(\brr) &= \left( x^2y^2z^2
  +{1\over 22}r^2\cdot\calY^{(A^+_1)}_{4}(\brr)-{1\over 105}r^6 \right)\;,\\
 \calY^{(A^+_1)}_{8}(\brr) &= \left(x^8+y^8+z^8-{28\over 5}r^2\cdot
 Y^{(A^+_1)}_{6}(\brr)
 -{210\over 143}r^4\cdot \calY^{(A^+_1)}_{4}(\brr)-{1\over 6}r^8\right)\;,\\
 ...
 \end{aligned} \right.
 \ee
 {\bf \HandRight\ The $A^-_2$ channel:}
 \be
 \label{eq:cubicY_A2}
 \left\{ \begin{aligned}
 \calY^{(A^-_2)}_{3}(\brr) &=xyz\;, \\
 \calY^{(A^-_2)}_{7}(\brr) &= xyz\left(x^4+y^4+z^4-{5\over 11} r^4\right)\;,\\
 ...
 \end{aligned} \right.
 \ee
 {\bf \HandRight\ The $E^+$ channel:}
 \be
 \label{eq:cubicY_E}
 \left\{ \begin{aligned}
 \calY^{(E^+)}_{2;1}(\brr) &=\left(z^2-{1\over 2}(x^2+y^2)\right)\;,
 \;
 \calY^{(E^+)}_{2;2}(\brr) =\left(x^2-y^2\right)\;,\\
 \calY^{(E^+)}_{4;1}(\brr) &= \left(z^4-{1\over 2}(x^4+y^4)
 -{6\over 7}r^2\cdot \calY^{(E^+)}_{2;1}(\brr)\right),
 \;
 \calY^{(E^+)}_{4;2}(\brr) = \left(x^4-y^4
 -{6\over 7}r^2\cdot \calY^{(E^+)}_{2;2}(\brr)\right)\;,\\
 \calY^{(E^+)}_{6;1}(\brr) &=\left(z^6-{1\over 2}(x^6+y^6)
 -{11\over 15}r^2\cdot \calY^{(E^+)}_{4;1}(\brr)
 -{5\over 7}r^4\cdot \calY^{(E^+)}_{2;1}(\brr)\right)\\
 \calY^{(E^+)}_{6;2}(\brr) &=\left(x^6-y^6
 -{11\over 15}r^2\cdot \calY^{(E^+)}_{4;2}(\brr)
 -{5\over 7}r^4\cdot \calY^{(E^+)}_{2;2}(\brr)\right)\;,\\
 ...
 \end{aligned} \right.
 \ee
 {\bf \HandRight\ The $T^-_1$ channel:}
 \be
 \label{eq:cubicY_T1}
 \left\{ \begin{aligned}
 \calY^{(T^-_1)}_{1;3}(\brr) &=\calY_{10}(\brr)\propto z\;,\\
 \calY^{(T^-_1)}_{3;3}(\brr) &=\calY_{30}(\brr)\propto z\left(z^2-{3\over 5}r^2\right)\;,\\
 \calY^{(T^-_1)}_{5;3}(\brr) &=z\left(z^4-{10\over 9}\left(z^2-{3\over 5}r^2\right)r^2
 -{3\over 7}r^4\right)\;,\\
 \calY^{(T^-_1)'}_{5;3}(\brr) &=z\left(x^4+y^4-{3\over
 4}(x^2+y^2)^2\right)\;,\\
 ...
 \end{aligned} \right.
 \ee
 {\bf \HandRight\ The $T^+_2$ channel:}
 \be
 \label{eq:cubicY_T2}
 \left\{ \begin{aligned}
 \calY^{(T^+_2)}_{2;3}(\brr) &=xy\;,\\
 \calY^{(T^+_2)}_{4;3}(\brr) &=xy\left(z^2-{1\over 7}r^2\right)\;,\\
 \calY^{(T^+_2)}_{6;3}(\brr)  &=xy\left(z^4-{6\over 11}z^2r^2+{1\over 33}r^4\right)\;,\\
 \calY^{(T^+_2)'}_{6;3}(\brr) &=xy\left(x^4+y^4-{5\over 8}(x^2+y^2)^2\right)\;,\\
 ...
 \end{aligned} \right.
 \ee
 \end{widetext}

 Note that for the irrep $T^-_1$ and $T^+_2$,
 the basis functions is formed by
 $3$ functions. Here we only list the ones that correspond to
 the $z$ components. Other two sets can be obtained by cyclicly
 swapping the coordinates.
 It is also noted that at $l=5$, we have two sets of basis (one of the basis function
 is denoted by a prime) since
 the irrep $T^-_1$ appears twice in the decomposition of
 $l=5$. Similar situation occurs for irrep $T^+_2$ at $l=6$.
 Since the function $G(\brr;k)$ is cubic invariant,
 it is easy to verify that $\calG^{(\Gamma)}_{l;i}$
 defined in this Appendix indeed forms a basis for the corresponding
 representation $\Gamma$ of the cubic group.



\begin{thebibliography}{6}
\providecommand{\natexlab}[1]{#1}
\providecommand{\url}[1]{\texttt{#1}}
\expandafter\ifx\csname urlstyle\endcsname\relax
  \providecommand{\doi}[1]{doi: #1}\else
  \providecommand{\doi}{doi: \begingroup \urlstyle{rm}\Url}\fi

\bibitem[Mathur et~al.(2004)Mathur, Lee, Alexandru, Bennhold, Chen, Dong,
  Draper, Horvath, Liu, Tamhankar, and Zhang]{KFLiu04:pentaquark}
N.~Mathur, F.X. Lee, A.~Alexandru, C.~Bennhold, Y.~Chen, S.J. Dong, T.~Draper,
  I.~Horvath, K.F. Liu, S.~Tamhankar, and J.B. Zhang.
\newblock A study of pentaquarks on the lattice with overlap fermions.
\newblock \emph{Phys. Rev. D}, 70:\penalty0 074508, 2004.

\bibitem[Niu et~al.(2009)Niu, Gong, Liu, and Shen]{Niu:2009gt}
Zhi-Yuan Niu, Ming Gong, Chuan Liu, and Yan Shen.
\newblock {Volume dependence of spectral weight functions}.
\newblock \emph{Phys. Rev.}, D80:\penalty0 114509, 2009.

\bibitem[L{\"u}scher(1991)]{luscher91:finitea}
M.~L{\"u}scher.
\newblock Two particle states on a torus and their relation to the scattering
  matrix.
\newblock \emph{Nucl. Phys. B}, 354:\penalty0 531, 1991.

\bibitem[Meng and Liu(2008)]{chuan08:Lee_model}
Guozhan Meng and Chuan Liu.
\newblock {Volume dependence of spectral weights for unstable particles in a
  solvable model}.
\newblock \emph{Phys. Rev.}, D78:\penalty0 074506, 2008.

\bibitem[von~der Lage and Bethe(1947)]{vonderLage:1947zz}
Fred~C. von~der Lage and H.~A. Bethe.
\newblock {A Method for Obtaining Electronic Eigenfunctions and Eigenvalues in
  Solids with An Application to Sodium}.
\newblock \emph{Phys. Rev.}, 71:\penalty0 612--622, 1947.

\bibitem[Landan and Lifshitz(1977)]{landau:quantum_mechanics_book}
L.D. Landan and E.M. Lifshitz.
\newblock \emph{Quantum Mechanics, 3rd ed.}
\newblock Pergamon Press, Oxford, UK, 1977.

\end{thebibliography}

\end{document}